\documentclass[12pt,a4paper]{article}

\usepackage{ifthen} 
\newboolean{pdflatex}
\setboolean{pdflatex}{false} 

\newboolean{articletitles}
\setboolean{articletitles}{true} 

\newboolean{uprightparticles}
\setboolean{uprightparticles}{false} 

\usepackage{lineno}  
\usepackage{xspace} 

\usepackage{graphicx}  
\usepackage{color}
\usepackage{colortbl}
\graphicspath{{./figs/}} 

\usepackage{amsmath} 
\usepackage{amssymb}
\usepackage{amsfonts}
\usepackage{upgreek} 

\newcommand*\patchAmsMathEnvironmentForLineno[1]{%
\expandafter\let\csname old#1\expandafter\endcsname\csname #1\endcsname
\expandafter\let\csname oldend#1\expandafter\endcsname\csname
end#1\endcsname
 \renewenvironment{#1}%
   {\linenomath\csname old#1\endcsname}%
   {\csname oldend#1\endcsname\endlinenomath}%
}
\newcommand*\patchBothAmsMathEnvironmentsForLineno[1]{%
  \patchAmsMathEnvironmentForLineno{#1}%
  \patchAmsMathEnvironmentForLineno{#1*}%
}
\AtBeginDocument{%
\patchBothAmsMathEnvironmentsForLineno{equation}%
\patchBothAmsMathEnvironmentsForLineno{align}%
\patchBothAmsMathEnvironmentsForLineno{flalign}%
\patchBothAmsMathEnvironmentsForLineno{alignat}%
\patchBothAmsMathEnvironmentsForLineno{gather}%
\patchBothAmsMathEnvironmentsForLineno{multline}%
}

\usepackage{hyperref}    
\usepackage[all]{hypcap} 

\def\lhcb {\mbox{LHCb}\xspace}
\def\ux85 {\mbox{UX85}\xspace}

\ifthenelse{\boolean{uprightparticles}}%
{

 \def\Ppi         {\ensuremath{\uppi}\xspace}

 \def\PDelta      {\ensuremath{\Delta}\xspace}                 
 \def\PXi      {\ensuremath{\Xi}\xspace}                 
 \def\PLambda      {\ensuremath{\Lambda}\xspace}                 
 \def\PSigma      {\ensuremath{\Sigma}\xspace}                 
 \def\POmega      {\ensuremath{\Omega}\xspace}                 
 \def\PUpsilon      {\ensuremath{\Upsilon}\xspace}                 
 
 \def\PB      {\ensuremath{\mathrm{B}}\xspace}                 
                  
 \def\PD      {\ensuremath{\mathrm{D}}\xspace}

 \def\PK      {\ensuremath{\mathrm{K}}\xspace}

 \def\Pb      {\ensuremath{\mathrm{b}}\xspace}                 
 \def\Pc      {\ensuremath{\mathrm{c}}\xspace}

 \def\Pi      {\ensuremath{\mathrm{i}}\xspace}

 \def\Ps      {\ensuremath{\mathrm{s}}\xspace}

}
{

 \def\Ppi         {\ensuremath{\pi}\xspace}

 \mathchardef\PDelta="7101
 \mathchardef\PXi="7104
 \mathchardef\PLambda="7103
 \mathchardef\PSigma="7106
 \mathchardef\POmega="710A
 \mathchardef\PUpsilon="7107
                  
 \def\PB      {\ensuremath{B}\xspace}                 
                  
 \def\PD      {\ensuremath{D}\xspace}

 \def\PK      {\ensuremath{K}\xspace}

 \def\Pb      {\ensuremath{b}\xspace}                 
 \def\Pc      {\ensuremath{c}\xspace}

 \def\Pi      {\ensuremath{i}\xspace}

 \def\Ps      {\ensuremath{s}\xspace}

}

\def\squark    {\ensuremath{\Ps}\xspace}

\def\cquark    {\ensuremath{\Pc}\xspace}

\def\bquark    {\ensuremath{\Pb}\xspace}

\def\pion  {\ensuremath{\Ppi}\xspace}

\def\pip   {\ensuremath{\pion^+}\xspace}
\def\pim   {\ensuremath{\pion^-}\xspace}

\def\kaon  {\ensuremath{\PK}\xspace}
  \def\Kbar  {\kern 0.2em\overline{\kern -0.2em \PK}{}\xspace}

\def\Kz    {\ensuremath{\kaon^0}\xspace}
\def\Kzb   {\ensuremath{\Kbar^0}\xspace}
\def\KzKzb {\ensuremath{\Kz \kern -0.16em \Kzb}\xspace}
\def\Kp    {\ensuremath{\kaon^+}\xspace}
\def\Km    {\ensuremath{\kaon^-}\xspace}

\def\KpKm  {\ensuremath{\Kp \kern -0.16em \Km}\xspace}

\def\Kstarz  {\ensuremath{\kaon^{*0}}\xspace}
\def\Kstarzb {\ensuremath{\Kbar^{*0}}\xspace}
\def\Kstarzbff {\ensuremath{\Kbar^{*}_{0}(1430)^{0}}\xspace}

  \def\Dbar    {\kern 0.2em\overline{\kern -0.2em \PD}{}\xspace}
\def\D       {\ensuremath{\PD}\xspace}

\def\Dz      {\ensuremath{\D^0}\xspace}
\def\Dzb     {\ensuremath{\Dbar^0}\xspace}
\def\DzDzb   {\ensuremath{\Dz {\kern -0.16em \Dzb}}\xspace}
\def\Dp      {\ensuremath{\D^+}\xspace}
\def\Dm      {\ensuremath{\D^-}\xspace}

\def\DpDm    {\ensuremath{\Dp {\kern -0.16em \Dm}}\xspace}

\def\Dstarp  {\ensuremath{\D^{*+}}\xspace}

\def\Ds      {\ensuremath{\D^+_\squark}\xspace}
\def\Dsp     {\ensuremath{\D^+_\squark}\xspace}
\def\Dsm     {\ensuremath{\D^-_\squark}\xspace}

\def\Dss     {\ensuremath{\D^{*+}_\squark}\xspace}
\def\Dssp    {\ensuremath{\D^{*+}_\squark}\xspace}

\def\B       {\ensuremath{\PB}\xspace}
  \def\Bbar    {\kern 0.18em\overline{\kern -0.18em \PB}{}\xspace}
\def\Bb      {\ensuremath{\Bbar}\xspace}
 
\def\Bz      {\ensuremath{\B^0}\xspace}
\def\Bzb     {\ensuremath{\Bbar^0}\xspace}

\def\Bub     {\ensuremath{\B^-}\xspace}

\def\Bm      {\ensuremath{\Bub}\xspace}

\def\Bs      {\ensuremath{\B^0_\squark}\xspace}
\def\Bsb     {\ensuremath{\Bbar^0_\squark}\xspace}

  \def\Y#1S{\ensuremath{\PUpsilon{(#1S)}}\xspace}

\def\Lbar {\ensuremath{\kern 0.1em\overline{\kern -0.1em\PLambda}}\xspace}

\def\Lc      {\ensuremath{\L^+_\cquark}\xspace}

\def\to                 {\ensuremath{\rightarrow}\xspace}

\def\AT#1     {\ensuremath{A_{\mathrm{T}}^{#1}}\xspace}           

\def\C#1      {\ensuremath{\mathcal{C}_{#1}}\xspace}                       
\def\Cp#1     {\ensuremath{\mathcal{C}_{#1}^{'}}\xspace}                    
\def\Ceff#1   {\ensuremath{\mathcal{C}_{#1}^{\mathrm{(eff)}}}\xspace}        
\def\Cpeff#1  {\ensuremath{\mathcal{C}_{#1}^{'\mathrm{(eff)}}}\xspace}       
\def\Ope#1    {\ensuremath{\mathcal{O}_{#1}}\xspace}                       
\def\Opep#1   {\ensuremath{\mathcal{O}_{#1}^{'}}\xspace}                    

\newcommand{\tev}{\ensuremath{\mathrm{\,Te\kern -0.1em V}}\xspace}
\newcommand{\gev}{\ensuremath{\mathrm{\,Ge\kern -0.1em V}}\xspace}
\newcommand{\mev}{\ensuremath{\mathrm{\,Me\kern -0.1em V}}\xspace}
\newcommand{\kev}{\ensuremath{\mathrm{\,ke\kern -0.1em V}}\xspace}
\newcommand{\ev}{\ensuremath{\mathrm{\,e\kern -0.1em V}}\xspace}
\newcommand{\gevc}{\ensuremath{{\mathrm{\,Ge\kern -0.1em V\!/}c}}\xspace}
\newcommand{\mevc}{\ensuremath{{\mathrm{\,Me\kern -0.1em V\!/}c}}\xspace}
\newcommand{\gevcc}{\ensuremath{{\mathrm{\,Ge\kern -0.1em V\!/}c^2}}\xspace}
\newcommand{\gevgevcccc}{\ensuremath{{\mathrm{\,Ge\kern -0.1em V^2\!/}c^4}}\xspace}
\newcommand{\mevcc}{\ensuremath{{\mathrm{\,Me\kern -0.1em V\!/}c^2}}\xspace}

\def\mum  {\ensuremath{\,\upmu\rm m}\xspace}

\newcommand{\chisq}{\ensuremath{\chi^2}\xspace}

\def\gsim{{~\raise.15em\hbox{$>$}\kern-.85em
          \lower.35em\hbox{$\sim$}~}\xspace}
\def\lsim{{~\raise.15em\hbox{$<$}\kern-.85em
          \lower.35em\hbox{$\sim$}~}\xspace}

\def\sWeights{\mbox{\em sWeights}}
\def\sWeight{\mbox{\em sWeight}}

\def\pt         {\mbox{$p_{\rm T}$}\xspace}

\def\evtgen     {\mbox{\textsc{EvtGen}}\xspace}
\def\pythia     {\mbox{\textsc{Pythia}}\xspace}

\def\geant      {\mbox{\textsc{Geant4}}\xspace}

\def\photos     {\mbox{\textsc{Photos}}\xspace}

\def\tell1  {TELL1\xspace}
\def\ukl1   {UKL1\xspace}

\def\dzdzb{{~\raise.85em\hbox{{\tiny{(}\textemdash\tiny{)}}}\kern-1.05em
          \lower0.0em\hbox{$D^0$}~}\xspace}
\def\bsbsb{{~\raise.85em\hbox{{\tiny{(}\textemdash\tiny{)}}}\kern-1.05em
          \lower0.0em\hbox{$B_s^0$}~}\xspace}

\def\br{{\cal{B}}}

\def\Lc{\Lambda_c^+}

\def\bstodspipipi{\Bsb\to\Dsp\pi^-\pi^+\pi^-}
\def\bstodskpipi{\Bsb\to\Dsp K^-\pi^+\pi^-}

\def\btodkpipi{\Bzb\to\Dp K^-\pi^+\pi^-}
\def\btodskpipi{\Bzb\to\Dsp K^-\pi^+\pi^-}
\def\btodsk{\Bzb\to\Dsp K^-}

\def\btodzerokpipi{\Bm\to D\Km\pip\pim}

\def\bstodsstarpipipi{\Bsb\to\Dss\pi^-\pi^+\pi^-}
\def\bstodsstarkpipi{\Bsb\to\Dss K^-\pi^+\pi^-}

\def\bstodsk{B_s^0\to D_s^{\mp}K^{\pm}}
\def\bstodskcc{\Bsb\to\Dsp\Km}

\def\btodzerok{\Bm\to D\Km}

\def\eff{\epsilon}

\def\ifb{\rm fb^{-1}}

\def\erel{\epsilon_{\rm rel}}
\def\eff{\epsilon}

\usepackage{mciteplus}

\begin{document}

\renewcommand{\thefootnote}{\fnsymbol{footnote}}
\setcounter{footnote}{1}


\begin{titlepage}
\pagenumbering{roman}

\vspace*{-1.5cm}
\centerline{\large EUROPEAN ORGANIZATION FOR NUCLEAR RESEARCH (CERN)}
\vspace*{1.5cm}
\hspace*{-0.5cm}
\begin{tabular*}{\linewidth}{lc@{\extracolsep{\fill}}r}
\ifthenelse{\boolean{pdflatex}}
{\vspace*{-2.7cm}\mbox{\!\!\!\includegraphics[width=.14\textwidth]{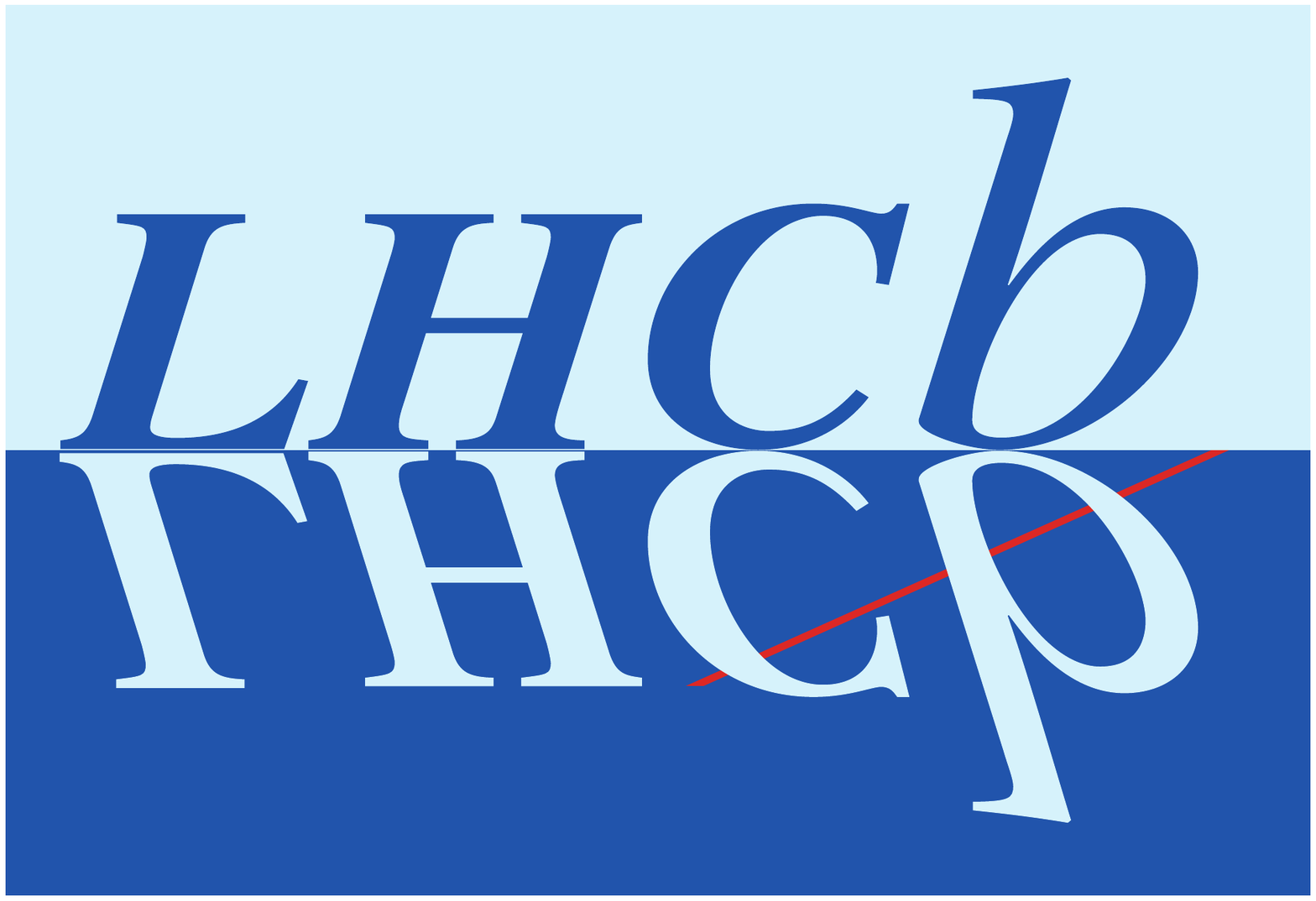}} & &}%
{\vspace*{-1.2cm}\mbox{\!\!\!\includegraphics[width=.12\textwidth]{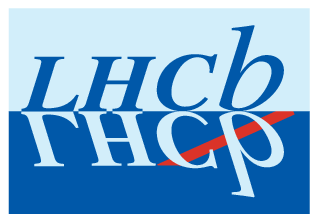}} & &}%
\\
 & & CERN-PH-EP-2012-327 \\  
 & & LHCb-PAPER-2012-033 \\  
 & & Dec. 24, 2012 \\ 
 & & \\
\end{tabular*}

\vspace*{0.5cm}

{\bf\boldmath\huge
\begin{center}
First observation of the
decays $\Bzb_{(s)}\to\Dsp\Km\pip\pim$ and $\Bsb\to D_{s1}(2536)^+\pim$
\end{center}
}

\vspace*{0.5cm}

\begin{center}
The LHCb collaboration\footnote{Authors are listed on the following pages.}
\end{center}

\vspace*{0.5cm}

\begin{abstract}
  \noindent
The first observation of the decays $\bstodskpipi$ and $\btodskpipi$ are reported using an 
integrated luminosity of 1.0~$\ifb$ recorded by the LHCb experiment. The branching fractions,
normalized with respect to $\bstodspipipi$ and $\bstodskpipi$, respectively, are measured to be
\begin{align*}
{\br(\bstodskpipi)\over\br(\bstodspipipi)} &= (5.2\pm0.5\pm0.3)\times10^{-2}, \nonumber \\
{\br(\btodskpipi)\over\br(\bstodskpipi)} &= 0.54\pm0.07\pm0.07, 
\end{align*}
\noindent where the first uncertainty is statistical and the second is systematic.
The $\bstodskpipi$ decay is of particular interest as it can be used to measure the weak phase $\gamma$.
First observation of the $\Bsb\to D_{s1}(2536)^+\pim,~D_{s1}^+\to\Dsp\pim\pip$ decay is also presented, 
and its branching fraction relative to $\bstodspipipi$ is found to be
\begin{align*}
{\br(\Bsb\to D_{s1}(2536)^+\pim,~D_{s1}^+\to\Dsp\pim\pip)\over\br(\bstodspipipi)} &= (4.0\pm1.0\pm0.4)\times10^{-3}. \\
\end{align*}

\end{abstract}

\vspace*{0.25cm}

\begin{center}
  Submitted to Physical Review D
\end{center}

\vspace{\fill}

\end{titlepage}


\newpage
\setcounter{page}{2}
\mbox{~}
\newpage

\centerline{\large\bf LHCb collaboration}
\begin{flushleft}
\small
R.~Aaij$^{38}$, 
C.~Abellan~Beteta$^{33,n}$, 
A.~Adametz$^{11}$, 
B.~Adeva$^{34}$, 
M.~Adinolfi$^{43}$, 
C.~Adrover$^{6}$, 
A.~Affolder$^{49}$, 
Z.~Ajaltouni$^{5}$, 
J.~Albrecht$^{35}$, 
F.~Alessio$^{35}$
% ...

{\footnotesize \it
% ...
$ ^{5}$Clermont Universit\'{e}, Universit\'{e} Blaise Pascal, CNRS/IN2P3, LPC, Clermont-Ferrand, France\\
$ ^{6}$CPPM, Aix-Marseille Universit\'{e}, CNRS/IN2P3, Marseille, France\\
% ...
$ ^{11}$Physikalisches Institut, Ruprecht-Karls-Universit\"{a}t Heidelberg, Heidelberg, Germany\\
% ...
$ ^{33}$Universitat de Barcelona, Barcelona, Spain\\
$ ^{34}$Universidad de Santiago de Compostela, Santiago de Compostela, Spain\\
$ ^{35}$European Organization for Nuclear Research (CERN), Geneva, Switzerland\\
% ...
$ ^{38}$Nikhef National Institute for Subatomic Physics, Amsterdam, The Netherlands\\
% ...
$ ^{43}$H.H. Wills Physics Laboratory, University of Bristol, Bristol, United Kingdom\\
% ...
$ ^{49}$Oliver Lodge Laboratory, University of Liverpool, Liverpool, United Kingdom\\
% ...
$ ^{n}$LIFAELS, La Salle, Universitat Ramon Llull, Barcelona, Spain\\
% ...
}
\end{flushleft}

\cleardoublepage


\renewcommand{\thefootnote}{\arabic{footnote}}
\setcounter{footnote}{0}



\pagestyle{plain} 
\setcounter{page}{1}
\pagenumbering{arabic}


%

\section{Introduction}

In the Standard Model (SM), the amplitudes associated with flavor-changing processes depend on
four Cabibbo-Kobayashi-Maskawa (CKM)~\cite{Cabibbo:1963yz,Kobayashi:1973fv} matrix parameters. Contributions from physics beyond
the Standard Model (BSM) add coherently to these amplitudes, leading to potential deviations in 
rates and CP-violating asymmetries when compared to the SM contributions alone.
Since the SM does not predict the CKM parameters, it is important
to make precise measurements of their values in processes that are expected to be insensitive to
BSM contributions. Their values then provide a benchmark to which BSM-sensitive measurements can be compared.

\begin{figure}[t]
\begin{center}
\includegraphics[width=0.45\linewidth]{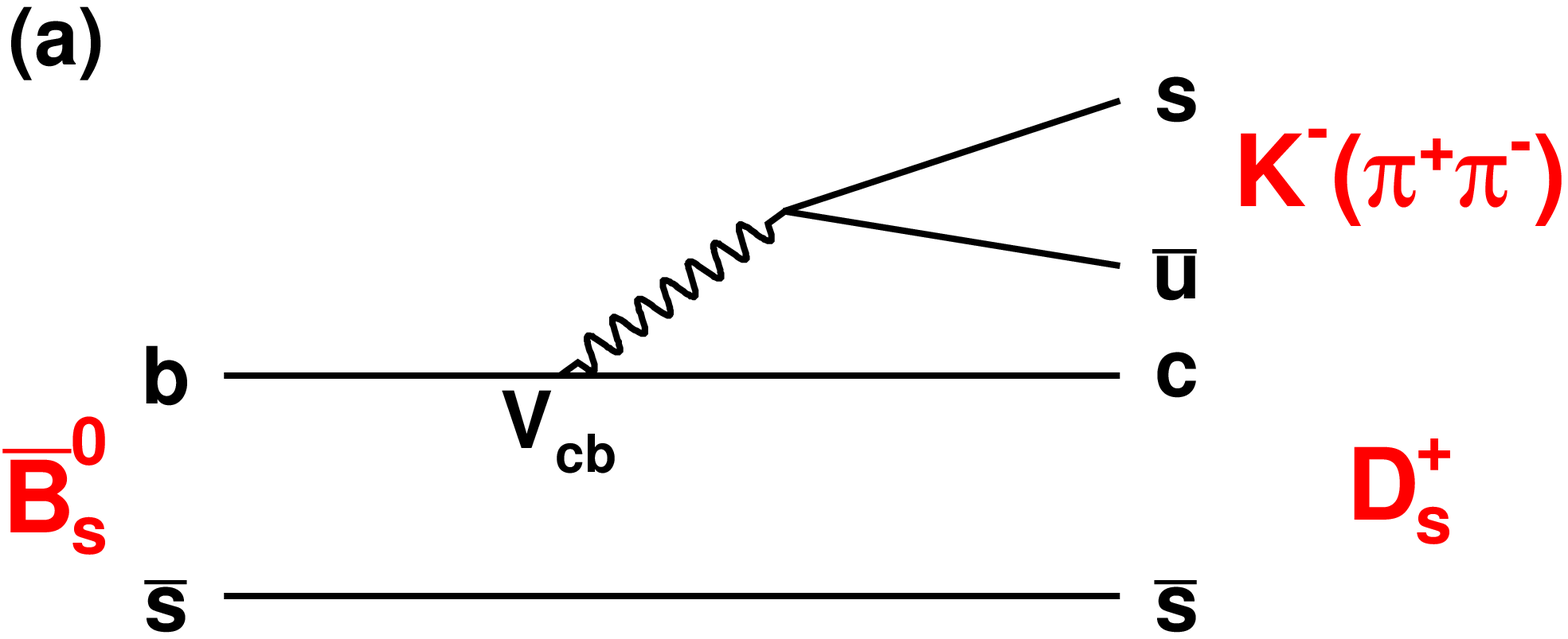}
\includegraphics[width=0.45\linewidth]{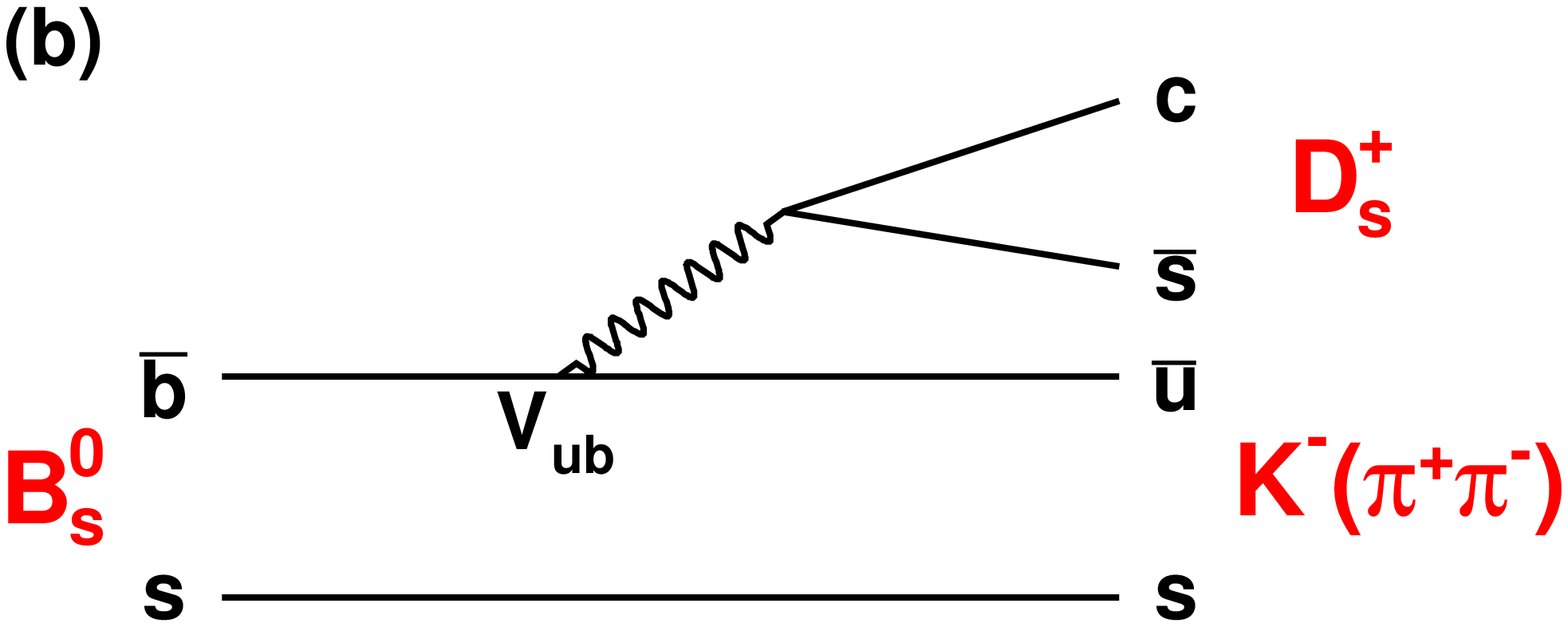}
\end{center}
\begin{center}
\includegraphics[width=0.45\linewidth]{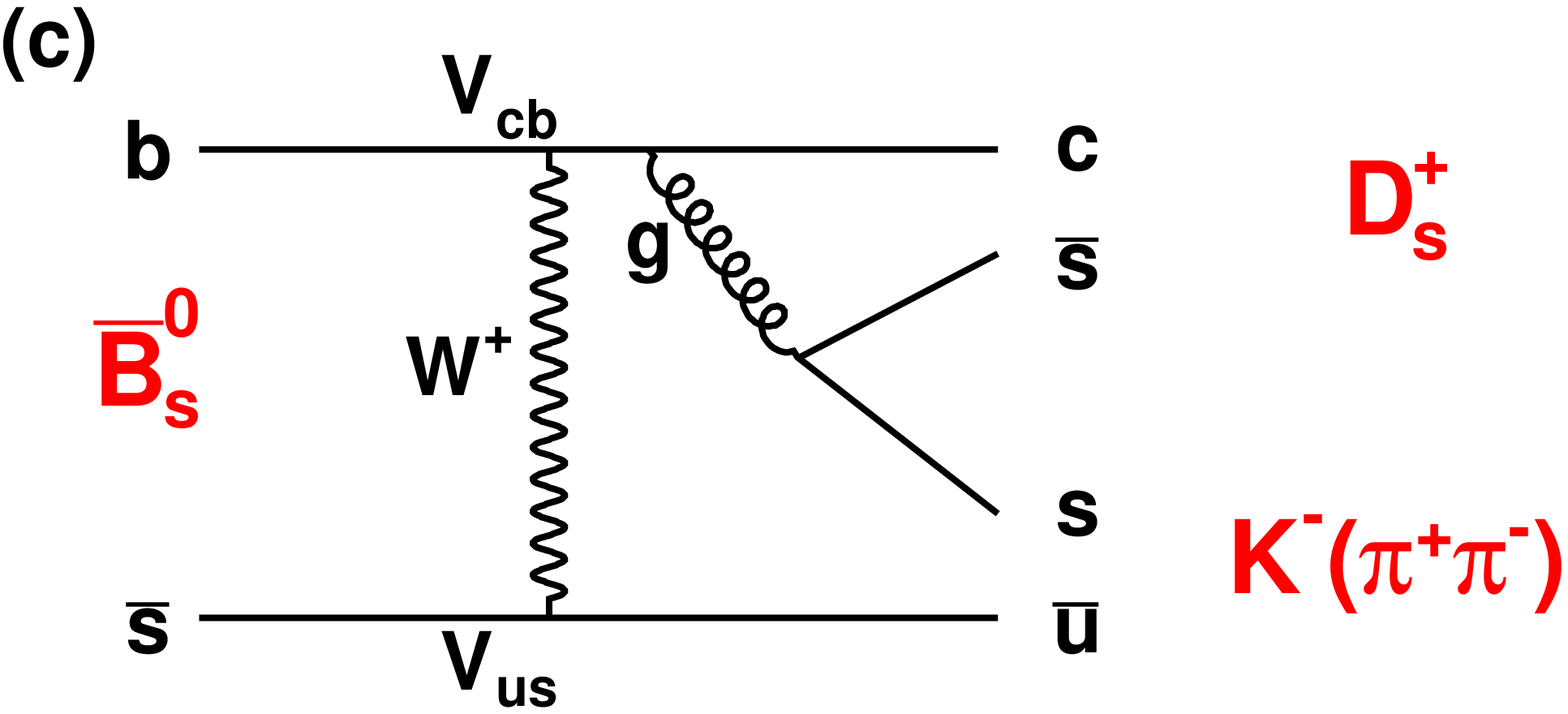}
\includegraphics[width=0.45\linewidth]{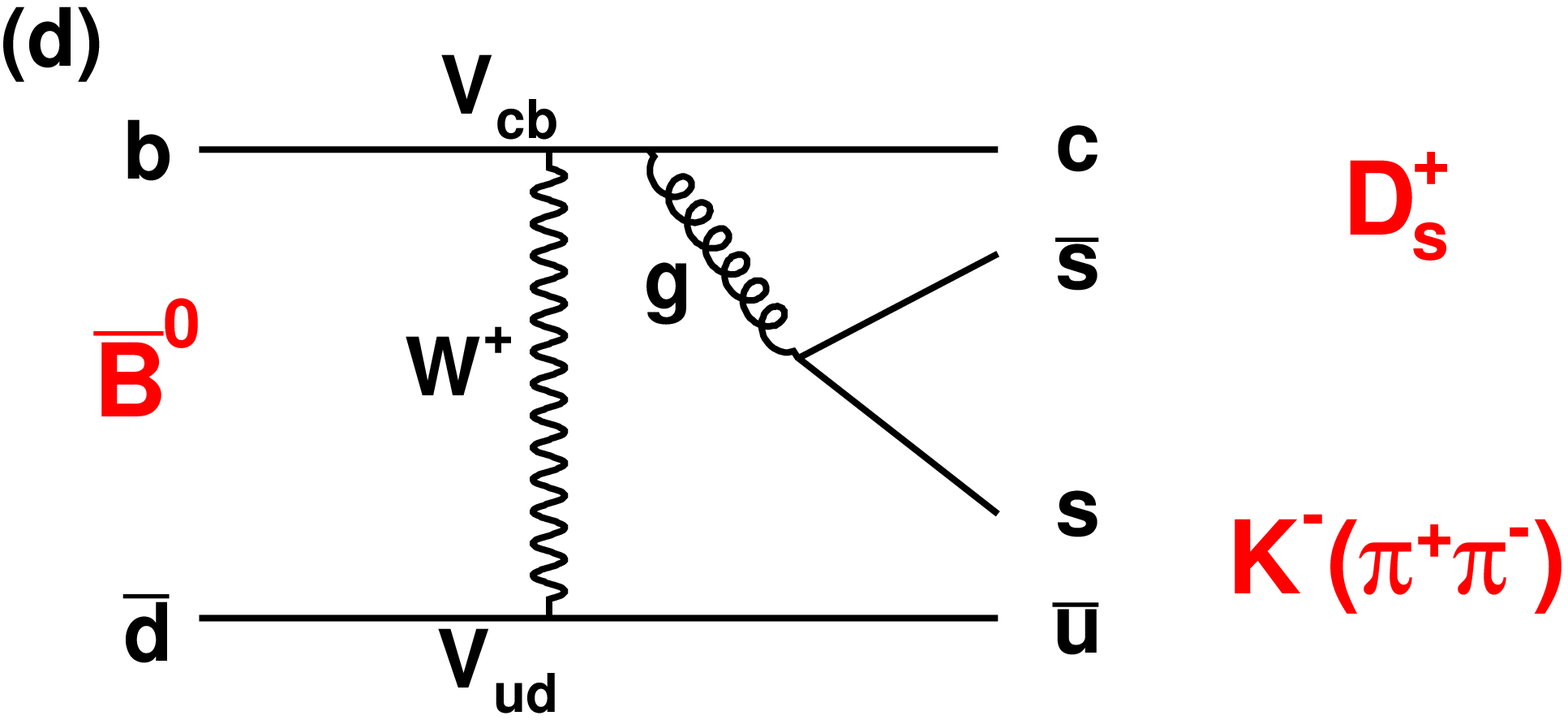}
\end{center}
\begin{center}
\includegraphics[width=0.45\linewidth]{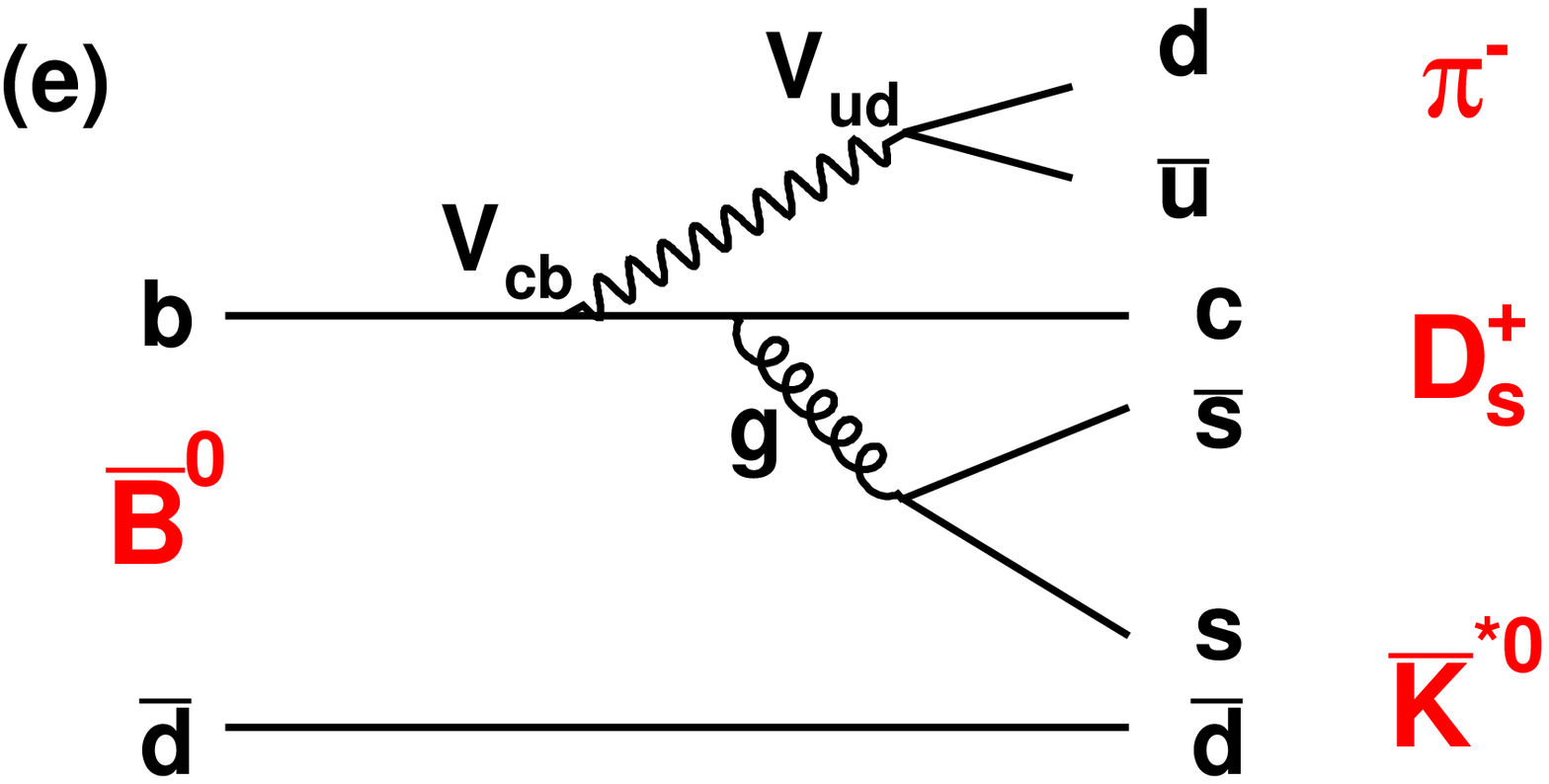}
\includegraphics[width=0.45\linewidth]{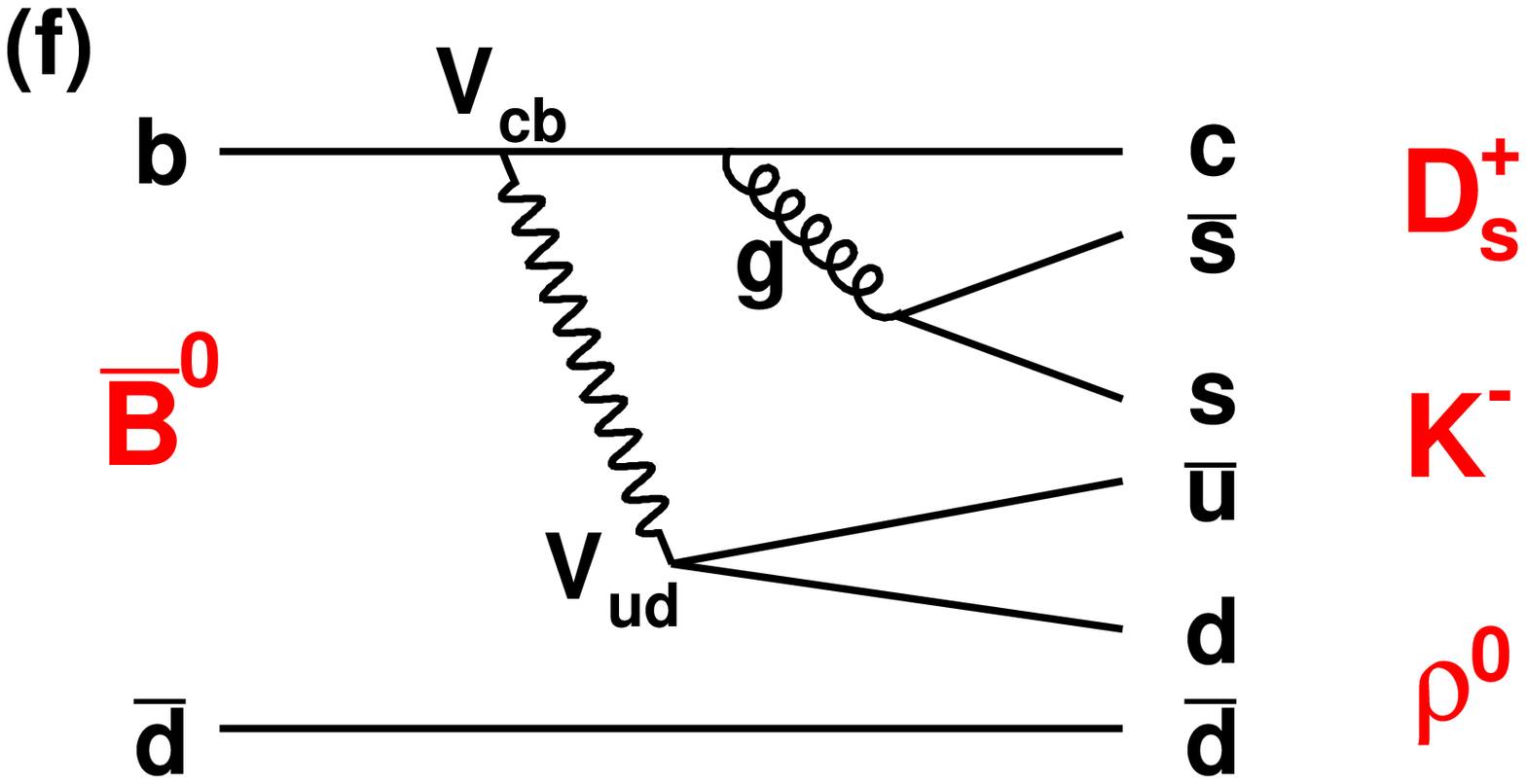}
\end{center}
\caption{Diagrams contributing to the $\Bs,\bstodskpipi$ (a-c) and $\btodskpipi$ (d-f) decays, as described in the text.
In (a-d), the additional ($\pip\pim$) indicates that the $\Km\pip\pim$ may be produced either through an excited strange kaon resonance
decay, or through fragmentation.}
\label{fig:feyn}
\end{figure}

The least well-determined of the CKM parameters is the weak phase 
${\gamma\equiv {\rm arg}\left(-{V_{\rm ub}^*V_{\rm ud}\over V_{\rm cb}^*V_{\rm cd}}\right)}$,
which, through direct measurements, is known to a precision of ${\sim10^{\rm o}-12^{\rm o}}$~\cite{Bona2012,Descotes2012}. 
It may be probed using time-independent 
rates of decays such as $\btodzerok$~\cite{Dunietz:1991yd,*Dunietz:1992ti,*Atwood:1994zm,*Atwood:1996ci,Gronau:1990ra,*Gronau:1991dp,Giri:2003ty},
or by analyzing the time-dependent decay rates of processes such as $\bstodsk$~\cite{Dunietz:1987bv,Aleksan:1991nh,Dunietz:1995cp,Fleischer:2003yb}.
Sensitivity to the weak phase $\gamma$ results from the interference between $b\to c$ and $b\to u$ transitions,
as indicated in Figs.~\ref{fig:feyn}(a-c). 
Such measurements may be extended to multibody decay modes, such as $\btodzerokpipi$~\cite{Aaij:2011rj} for a time-independent measurement,
or $\bstodskpipi$ in the case of a time-dependent analysis. 

The $\btodkpipi$ decay, while having the same final state as $\bstodskpipi$, receives contributions not only from the 
$W$-exchange process (Fig.~\ref{fig:feyn}(d)), but also from $b\to c$ transitions in association with the production of an extra 
$s\bar{s}$ pair (Figs.~\ref{fig:feyn}(e-f)). The decay may also proceed through mixing followed by a $b\to u$, $W$-exchange process (not shown). 
However, this amplitude is Cabibbo-, helicity- and color-suppressed, and is therefore negligible compared to the $b\to c$ amplitude.

This paper reports the first observation of $\bstodskpipi$ and $\btodskpipi$, and 
measurements of their branching fractions relative to $\bstodspipipi$ and $\bstodskpipi$, respectively.
The data sample is based on an integrated
luminosity of 1.0~$\ifb$ of $pp$ collisions at $\sqrt{s}=7$~\tev 
collected by the LHCb experiment. The same data sample is also used to observe the 
$\Bsb\to D_{s1}(2536)^+\pim,~D_{s1}^+\to\Dsp\pim\pip$ decay for the first time, 
and measure its branching fraction relative to $\bstodspipipi$.
The inclusion of charge-conjugated modes is implied throughout this paper.

\section{Detector and simulation}
\label{sec:Detector}

The \lhcb detector~\cite{Alves:2008zz} is a single-arm forward
spectrometer covering the \mbox{pseudorapidity} range $2<\eta <5$, designed
for the study of particles containing \bquark or \cquark quarks. The
detector includes a high precision tracking system consisting of a
silicon-strip vertex detector surrounding the $pp$ interaction region,
a large-area silicon-strip detector located upstream of a dipole
magnet with a bending power of about $4{\rm\,Tm}$, and three stations
of silicon-strip detectors and straw drift-tubes placed
downstream. The combined tracking system has a momentum resolution
($\Delta p/p$) that varies from 0.4\% at 5\gevc to 0.6\% at 100\gevc,
and an impact parameter (IP) resolution of 20\mum for tracks with high
transverse momentum ($\pt$). Charged hadrons are identified using two
ring-imaging Cherenkov (RICH) detectors. Photon, electron and hadron
candidates are identified by a calorimeter system consisting of
scintillating-pad and pre-shower detectors, an electromagnetic
calorimeter and a hadronic calorimeter. Muons are identified by a
system composed of alternating layers of iron and multiwire
proportional chambers. 

The trigger consists of a hardware stage, based
on information from the calorimeter and muon systems, followed by a
software stage, which applies a full event reconstruction.
The software trigger requires a two-, three- or four-track
secondary vertex with a high \pt sum of the tracks and a significant displacement from the 
primary $pp$ interaction vertices~(PVs).  At least one track should have $\pt > 1.7\gevc$,
an IP~\chisq greater than 16 with respect to all PVs, 
and a track fit $\chisq/\rm{ndf} < 2$, where ndf is the number of degrees of freedom. 
The IP \chisq is defined as the difference between the \chisq of the PV
reconstructed with and without the considered particle. A multivariate
algorithm is used for the identification of secondary
vertices~\cite{LHCb-PUB-2011-016}.

For the simulation, $pp$ collisions are generated using
\pythia~6.4~\cite{Sjostrand:2006za} with a specific \lhcb
configuration~\cite{LHCb-PROC-2010-056}.  Decays of hadronic particles
are described by \evtgen~\cite{Lange:2001uf} in which final state
radiation is generated using \photos~\cite{Golonka:2005pn}. The
interaction of the generated particles with the detector and its
response are implemented using the \geant
toolkit~\cite{Allison:2006ve, *Agostinelli:2002hh} as described in
Ref.~\cite{LHCb-PROC-2011-006}.

\section{Signal selection}

Signal $\Bzb_{(s)}$ decay candidates are formed by pairing a $\Dsp\to\Kp\Km\pip$ candidate with
either a $\pim\pip\pim$ (hereafter referred to as $X_d$) or a $\Km\pip\pim$ combination (hereafter referred to as $X_s$).
Tracks used to form the $\Dsp$ and $X_{d,s}$ are required to be identified as either a pion or
a kaon using information from the RICH detectors, have $\pt$ in excess of 100~\mevc, and be 
significantly detached from any reconstructed PV in the event.

Signal $\Dsp$ candidates are required to have good vertex fit quality, be significantly displaced from
the nearest PV, and have invariant mass, $M(\Kp\Km\pip)$, within 20~\mevcc of the $\Dsp$ 
mass~\cite{Beringer:1900zz}. To suppress combinatorial 
and charmless backgrounds, only those $\Dsp$ candidates that are consistent with decaying through either the
$\phi$ ($M(\Kp\Km)<1040~\mevcc$) or $\Kstarzb$ ($|M(\Km\pip)-m_{\Kstarz}|<75~\mevcc$) resonances are used 
(here, $m_{\Kstarz}$ is the $\Kstarz$ mass~\cite{Beringer:1900zz}). The remaining
charmless background yields are determined using the $\Dsp$ mass sidebands.
For about 20\% of candidates, when the $\Kp$ is assumed to be a $\pip$, the corresponding $\Km\pip\pip$ 
invariant mass is consistent with the $\Dp$ mass. To suppress cross-feed from $\Bzb\to\Dp X$ decays, 
a tighter particle identification (PID) requirement is applied to the $\Kp$ in the $\Ds\to\Kp\Km\pip$ candidates
when $|M(\Km\pip\pip)-m_{\Dp}|<20~\mevcc$ ($m_{\Dp}$ is the $\Dp$ mass~\cite{Beringer:1900zz}). 
Similarly, if the invariant mass of the particles forming the $\Dsp$ candidate, after replacing the 
$\Kp$ mass with the proton mass, falls within 15~\mevcc of the $\Lc$ mass, tighter PID selection is applied. 
The sizes of these mass windows are about 2.5 times the invariant mass resolution, and are sufficient to render these cross-feed
backgrounds negligible.

Candidate $X_d$ and $X_s$ are formed from $\pim\pip\pim$ or $\Km\pip\pim$ combinations, where all invariant mass values up to 
3~\gevcc are accepted.
To reduce the level of combinatorial background, we demand that the $X_{d,s}$ vertex is displaced from the nearest PV by
more than 100~\mum in the direction transverse to the beam axis and that at least two of the daughter tracks have 
$\pt>300~\mevc$. Backgrounds to the $\Bzb_{(s)}\to\Dsp\Km\pip\pim$ search from $\Bsb\to D_s^{(*)+}\pim\pip\pim$ or
$\Bsb\to\Dsp\Km\Kp\pim$ decays are suppressed by applying more stringent PID requirements to the $\Km$ and $\pip$ in $X_s$.
The PID requirements have an efficiency of about 65\% for 
selecting $X_s$, while rejecting about 97\% of the favored three-pion background.
To suppress peaking backgrounds from $\Bsb\to\Dsp\Dsm$ decays, where $\Dsp\to\pip\pim\pip,~\Kp\pim\pip$, 
it is required that $M(X_{d,s})$ is more than 20~\mevcc away from the $\Dsp$ mass.

Signal $\Bb$ meson candidates are then formed by combining a $\Dsp$ with either an $X_d$ or $X_s$. 
The reconstructed $\Bb$ candidate is
required to be well separated from the nearest PV with a decay time larger than 0.2~ps and have a good quality
vertex fit. To suppress remaining charmless backgrounds, which appear primarily in
$\btodskpipi$, the vertex separation (VS) $\chi^2$ between the $\Dsp$ and $\Bb$ decay vertices is required to 
be greater than 9. Candidates passing all selection requirements are refit with both $\Dsp$ mass and vertex constraints
to improve the mass resolution~\cite{Hulsbergen:2005pu}. 

To further suppress combinatorial background, a boosted decision tree (BDT) selection~\cite{Breiman,*Roe} with the AdaBoost 
algorithm\cite{AdaBoost} is employed. The BDT
is trained using simulated $\bstodskpipi$ decays for the signal distributions, and the high $\Bb$ mass sideband 
in data are used to model the backgrounds. The following thirteen variables are used:
\begin{itemize}
\item $\Bb$ candidate: {\rm IP}~\chisq, VS~\chisq, vertex fit $\chisq$, and $\pt$;
\item $\Dsp$ candidate: Flight distance significance from $\Bb$ vertex;
\item $X_{d,s}$ candidate: IP~\chisq, maximum of the distances of closest approach between any pair of tracks in the decay;
\item $X_{d,s}$ daughters: min(IP~\chisq), max(IP~\chisq), min(\pt);
\item $\Dsp$ daughters: min(IP~\chisq), max(IP~\chisq), min(\pt),
\end{itemize}
\noindent where min and max denote the minimum and maximum of the indicated values amongst the daughter particles.
The flight distance significance is the separation between the $\Dsp$ and $\Bb$ vertices, normalized by the uncertainty.
The training produces a single variable, $x$, that provides discrimination between
signal decays and background contributions. The cut value is chosen by optimizing $S(x_{\rm cut})/\sqrt{S(x_{\rm cut})+B(x_{\rm cut})}$, where
$S(x_{\rm cut})$ and $B(x_{\rm cut})$ are the expected signal and background yields, respectively, after
requiring $x>x_{\rm cut}$. At the optimal point, a signal efficiency of $\sim$90\% is expected while
rejecting about $85\%$ of the combinatorial background (after the previously discussed selections are applied). 
After all selections, about 3\% of events have
more than one signal candidate in both data and simulation. All candidates are kept for further analysis.

\section{Fits to data}

The $\bstodspipipi$ and $\Bzb_{(s)}\to\Dsp\Km\pip\pim$ invariant mass spectra are each modeled by the
sum of a signal and several background components. The signal shapes are obtained from simulation,
and are each described by the sum of a Crystal Ball (CB)~\cite{Skwarnicki:1986xj} shape and a Gaussian function.
The CB shape parameter that describes the tail toward low mass is fixed based on simulated decays. A common, 
freely varying scale factor multiplies the width parameters in the CB
and Gaussian functions to account for slightly larger resolution in data than in simulation. For the $\Bzb_{(s)}\to\Dsp\Km\pip\pim$ 
mass fit, the difference between the mean $\Bsb$ and $\Bzb$ masses is fixed to 87.35~\mevcc~\cite{Beringer:1900zz}.

Several non-signal $b$-hadron decays produce broad peaking structures in the $\Dsp\pim\pip\pim$ and 
$\Dsp\Km\pip\pim$ invariant mass spectra.
For $\bstodspipipi$, the only significant source of peaking background is from $\bstodsstarpipipi$, where the 
photon or $\pi^0$ from the $\Dssp$ decay is not included in the reconstructed decay. Since
the full decay amplitude for $\bstodsstarpipipi$ is not known, the simulation may not adequately model the
decay. Simulation is therefore used to provide an estimate for the shape, but the parameters are
allowed to vary within one standard deviation about the fitted values.

For $\Bzb_{(s)}\to\Dsp\Km\pip\pim$, backgrounds from $\Bzb_{(s)}\to\Dss\Km\pip\pim$
and from misidentified $\bstodspipipi$ and $\bstodsstarpipipi$ decays are considered. The $\Bzb_{(s)}\to\Dss\Km\pip\pim$
shape is fixed to be the same as that obtained for the $\bstodsstarpipipi$ component in the $\bstodspipipi$ mass fit.
This same shape is assumed for both $\Bzb$ and $\Bsb$, where for the former, 
a shift by the $\Bzb-\Bsb$ mass difference is included. For the $\bstodspipipi$ and $\bstodsstarpipipi$ cross-feed,
simulated decays and kaon misidentification rates taken from $\Dstarp$ calibration data are used to obtain their expected yields and
invariant mass shapes. The cross-feed contribution is about 3\% of the $\bstodspipipi$ and $\bstodsstarpipipi$ yields;
the corresponding cross-feed yields are fixed in the $\Bzb_{(s)}\to\Dsp\Km\pip\pim$ fit. The shape is obtained by parameterizing the invariant
mass spectrum obtained from the simulation after replacing the appropriate $\pim$ mass in $X_d$ with the kaon mass.
The combinatorial background is described by an exponential function whose slope is allowed to vary independently for both mass fits.

Figure~\ref{fig:MassPlot_B2DsPiPiPi} shows the invariant mass distribution for $\bstodspipipi$
candidates passing all selection criteria. The fitted number of $\bstodspipipi$ signal events is
$5683\pm83$. While it is expected that most
of the low mass background emanates from $\bstodsstarpipipi$ decays, contributions from other sources such
as $\Bsb\to\Ds\pim\pip\pim\pi^0$ are also possibly absorbed into this background component.
Figure~\ref{fig:MassPlot_B2DsKPiPi} shows the invariant mass distribution for $\Bzb_{(s)}\to\Dsp\Km\pip\pim$ 
candidates. 
The fitted signal yields are $402\pm33$ $\btodskpipi$ and $216\pm21$ $\bstodskpipi$ events.

\begin{figure}[t]
\begin{center}
\includegraphics[width=1.0\linewidth]{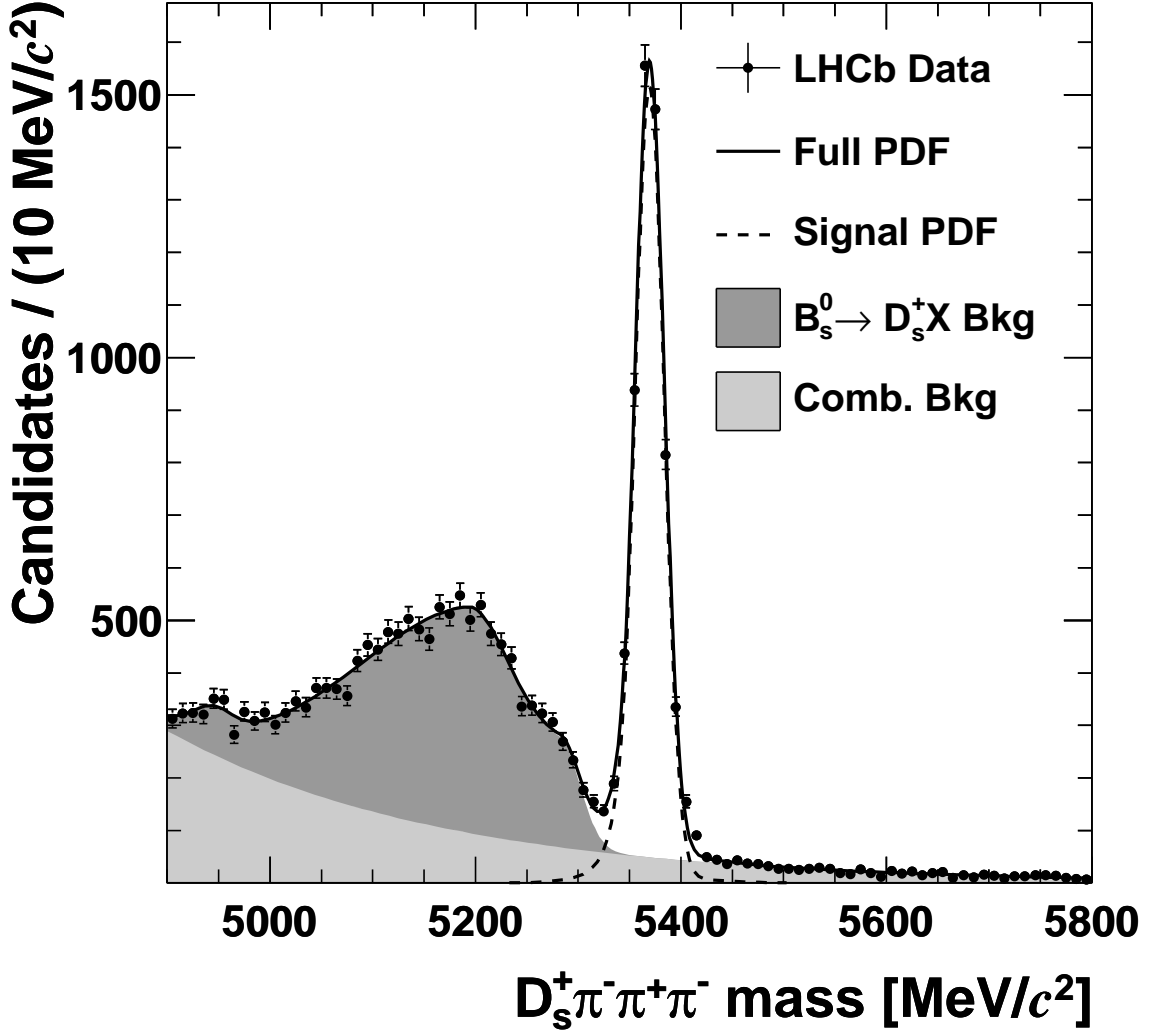}
\end{center}
\caption{Invariant mass distribution for $\bstodspipipi$ candidates.
The fitted signal probability disrtibution function (PDF) is indicated by the dashed line 
and the background shapes are shown as shaded regions, as described in the text.}
\label{fig:MassPlot_B2DsPiPiPi}
\end{figure}

\begin{figure}[t]
\begin{center}
\includegraphics[width=1.0\linewidth]{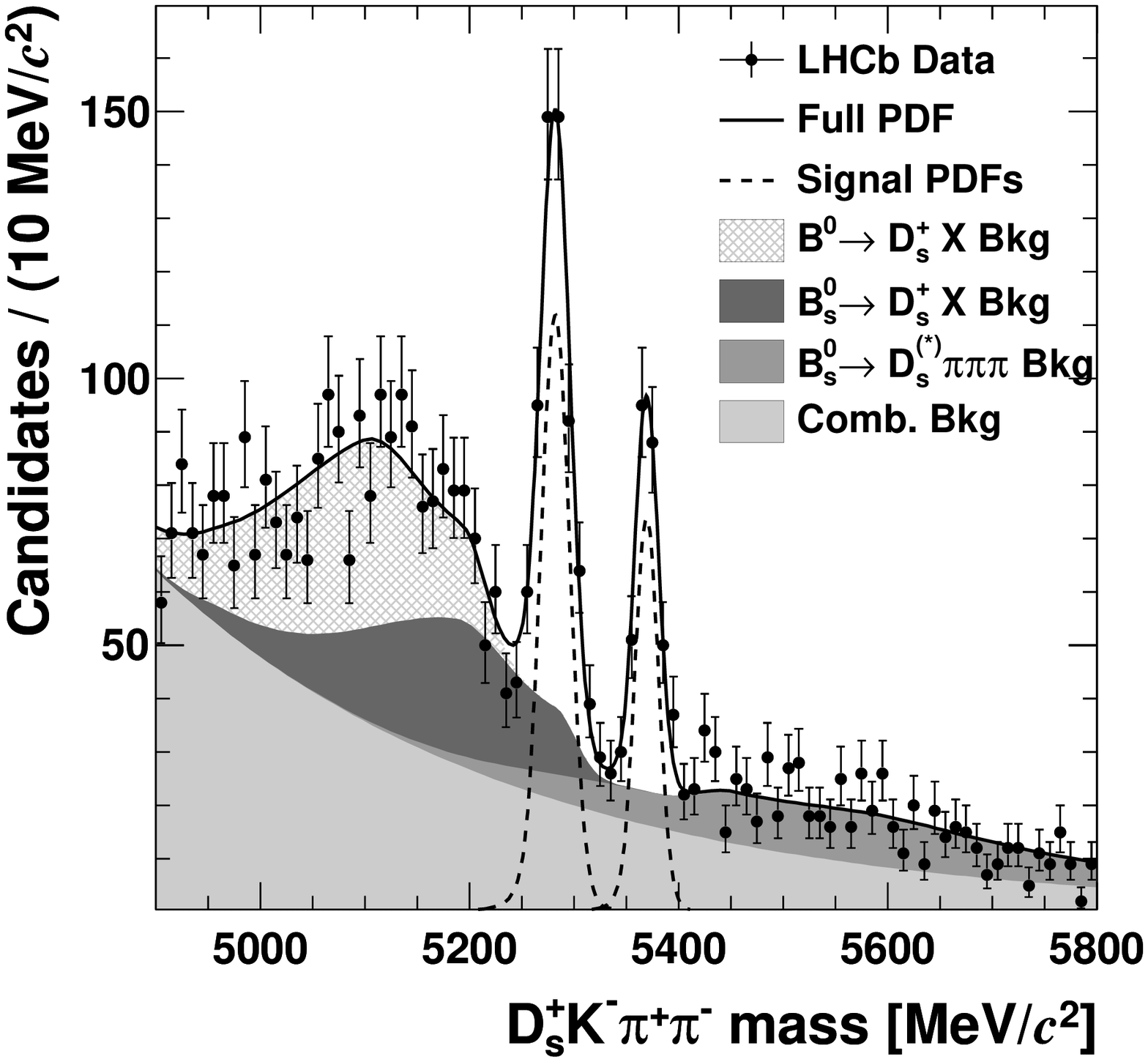}
\end{center}
\caption{Invariant mass distribution for $\Bzb_{(s)}\to\Dsp\Km\pip\pim$ candidates.
The fitted signal (dashed lines) and background shapes (shaded/hatched regions) are shown, as described in the text.}
\label{fig:MassPlot_B2DsKPiPi}
\end{figure}

The $\Dsp$ mass sidebands, defined to be from 35 to 55~\mevcc on either side of the nominal $\Dsp$ mass,
are used to estimate the residual charmless background that may 
contribute to the observed signals. The numbers of $\Bs$ decays in the $\Dsp$ sidebands are 
$61\pm16$ , $0^{+5}_{-0}$, and $9\pm5$ 
for the $\bstodspipipi$, $\bstodskpipi$ and $\btodskpipi$ decays, respectively; they
are subtracted from the observed signal yields to obtain the corrected number of signal decays. 
The yields in the signal and sideband regions are summarized in Table~\ref{tab:yields}.
\begin{table*}[h]
\begin{center}
\caption{Summary of event yields from data in the $\Dsp$ signal and sidebands regions, and the background corrected yield.
The signal and sideband regions require $\Dsp$ candidates to have invariant mass $|M(\Kp\Km\pip)-m_{\Dsp}|<20$~\mevcc
and $35<|M(\Kp\Km\pip)-m_{\Dsp}|<55$~\mevcc, respectively, where $m_{\Dsp}$ is the $\Dsp$ mass~\cite{Beringer:1900zz}.}
\begin{tabular}{lccc}
\hline\hline
Decay           & Signal Region  & Sideband Region & Corrected Yield \\
\hline \\[-2.20ex]
$\bstodspipipi$ & $5683\pm83$    & $\phantom{1}61\pm16$       & $5622\pm85$ \\
$\bstodskpipi$  & $\phantom{1}216\pm21$     & $\phantom{1}0^{+5}_{-0}$   & $\phantom{1}216\pm22$ \\
$\btodskpipi$   & $\phantom{1}402\pm33$     & $~~9\pm5$         & $\phantom{1}393\pm33$ \\
\hline\hline
\end{tabular}
\label{tab:yields}
\end{center}
\end{table*}

\section{Mass distributions of ${\mathbf X_{\boldsymbol{d,s}}}$ and two-body masses}
\label{sec:kinematics}

In order to investigate the properties of these $\Bzb_{(s)}$ decays, \sWeights~\cite{Pivk:2004ty} obtained from the mass fits are used to
determine the underlying $X_{d,s}$ invariant mass spectra as well as the two-body invariant masses amongst the three daughter 
particles. Figure~\ref{fig:Bs2Ds3Pi_Masses} shows (a) the $\pim\pip\pim$ mass, (b) the smaller $\pip\pim$ mass and (c) the larger
$\pip\pim$ mass in $\bstodspipipi$ data and simulated decays. A prominent peak, consistent with
the $a_1(1260)^-\to\pim\pip\pim$ is observed, along with structures consistent with the $\rho^0$ in the two-body
masses. There appears to be an offset in the peak position of the $a_1(1260)^-$ between data and
simulation. Since the mean and width of the $a_1(1260)^-$ resonance are not well known, and their values may even be 
process dependent, this level of agreement is reasonable. 
A number of other spectra have been compared between data and simulation, such as the \pt spectra of the
$\Dsp$, $X_d$ and the daughter particles, and excellent agreement is found.

\begin{figure}[t]
\begin{center}
\includegraphics[width=1.0\linewidth]{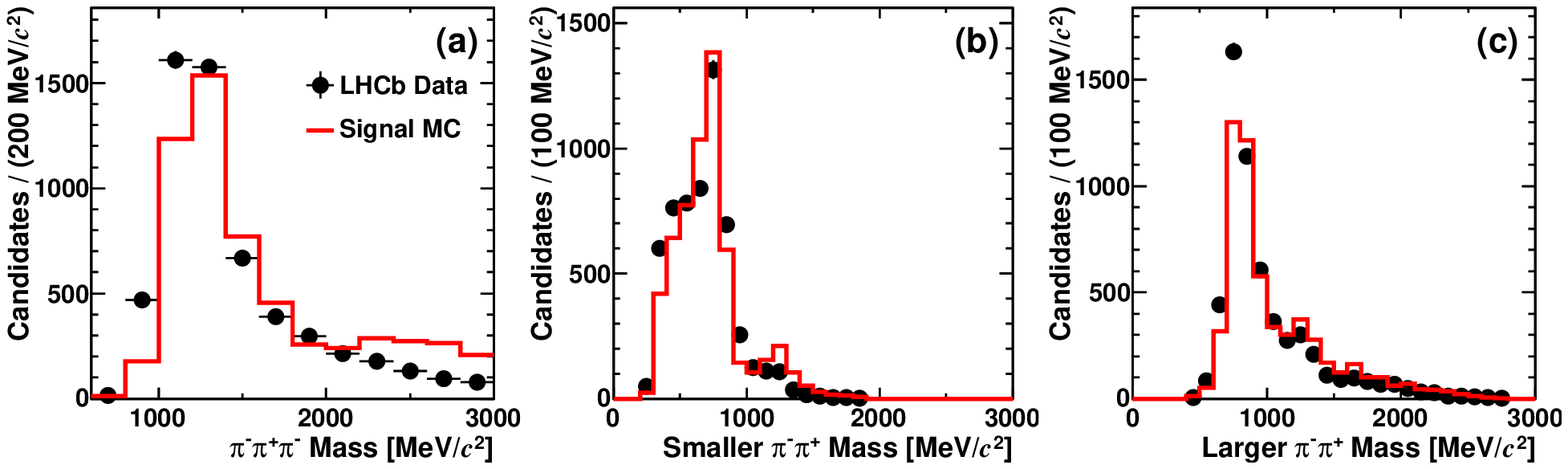}
\end{center}
\caption{Invariant mass distributions for (a) $X_d$, (b) smaller $\pip\pim$ mass in $X_d$ and (c)
the larger $\pip\pim$ mass in $X_d$, from $\bstodspipipi$ decays using \sWeights. 
The points are the data and the solid line is the simulation. The simulated distribution is 
normalized to have the same yield as the data.}
\label{fig:Bs2Ds3Pi_Masses}
\end{figure}

Figure~\ref{fig:Bs2DsK2Pi_Masses} shows the corresponding distributions for the $\bstodskpipi$ decay. A peaked structure
at low $\Km\pip\pim$ mass, consistent with contributions from the lower-lying excited strange mesons, such
as the $K_1(1270)^-$ and $K_1(1400)^-$, is observed. As many of these states
decay through $\Kstarzb$ and $\rho^0$ mesons, significant contributions from these resonances are observed in the $\Km\pip$ and 
$\pip\pim$ invariant mass spectra, respectively. The simulation provides a reasonable description of the distributions in the data.

\begin{figure}[t]
\begin{center}
\includegraphics[width=1.0\linewidth]{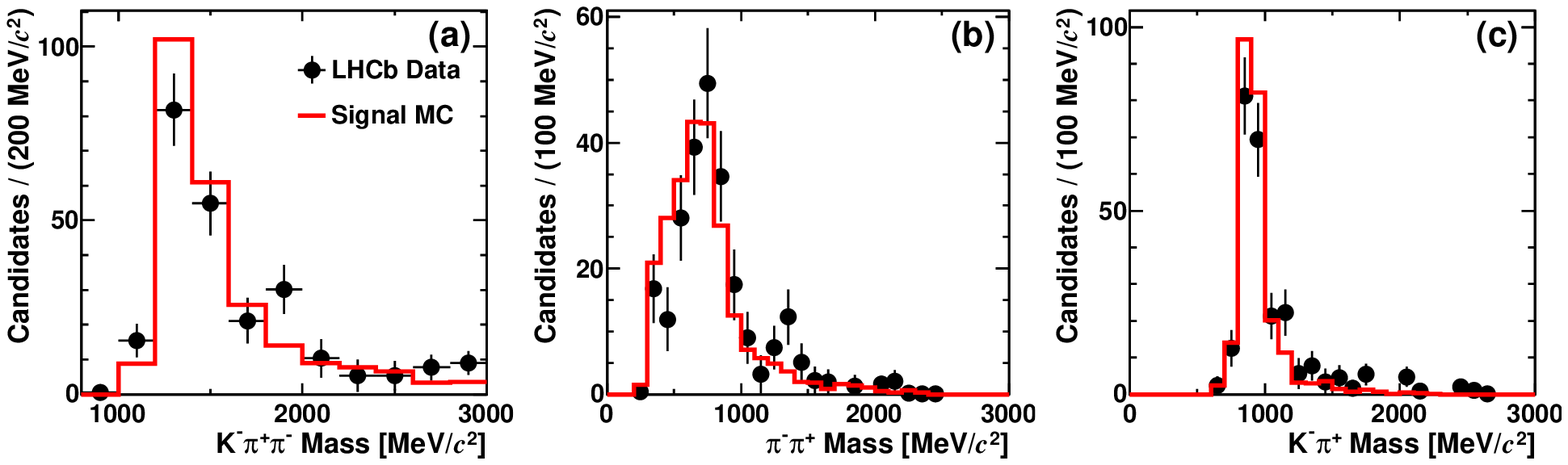}
\end{center}
\caption{Invariant mass distributions for (a) $X_s$, (b) $\pip\pim$ in $X_s$ and (c)
the $\Km\pip$ in $X_s$, from $\bstodskpipi$ data using \sWeights. The points are data and the solid line is the simulation.
The simulated distribution is normalized to have the same yield as the data.}
\label{fig:Bs2DsK2Pi_Masses}
\end{figure}

Figure~\ref{fig:Bd2DsK2Pi_Masses} shows the same distributions for $\btodskpipi$. 
The $\Km\pip\pim$ invariant mass is quite broad, with little indication of any narrow structures. There are indications
of $\Kstarzb$ and $\rho^0$ contributions in the $\Km\pip$ and $\pim\pip$ invariant mass spectra, respectively, but the contribution
from resonances such as the  $K_1(1270)^-$ or $K_1(1400)^-$ appear to be small or absent.
In the $\Km\pip$ invariant mass spectrum, there may be an indication of a $\Kstarzbff$ contribution.
The simulation, which models the $\Km\pip\pim$ final state as 10\% $K_1(1270)^-$, 10\% $K_1(1400)^-$, 40\% $\Kstarzb\pim$ and
40\% $\Km\rho^0$, provides a reasonable description of the data, which suggests that processes such as those in
Figs.~\ref{fig:feyn}(e-f) constitute a large portion of the total width for this decay.

\begin{figure}[tb]
\begin{center}
\includegraphics[width=1.0\linewidth]{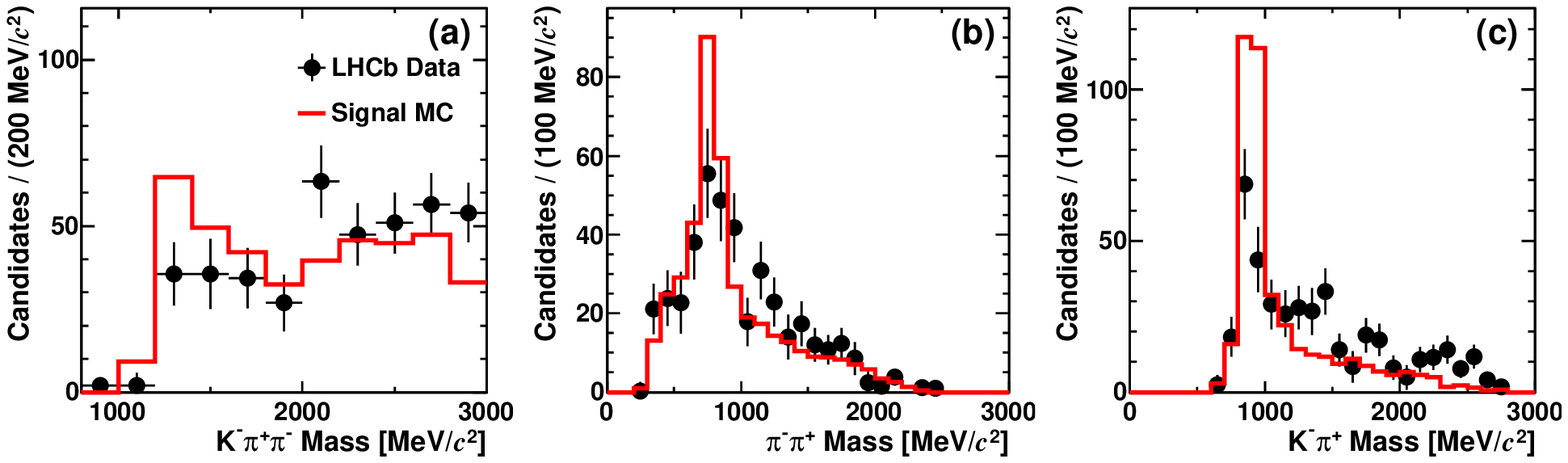}
\end{center}
\caption{Invariant mass distributions for (a) $X_s$, (b) $\pip\pim$ in $X_s$ and (c)
the $\Km\pip$ in $X_s$, from  $\btodskpipi$ data using \sWeights. The points are data and the solid line is the simulation.
The simulated distribution is normalized to have the same yield as the data.}
\label{fig:Bd2DsK2Pi_Masses}
\end{figure}

\section{First observation of $\boldsymbol{\Bsb\to~D_{s1}(2536)^+\pim}$ }

A search for excited $\Ds$ states, such as $D_{sJ}^+\to\Dsp\pim\pip$, contributing to the $\bstodspipipi$ final state
is performed. Signal candidates within $\pm$40~\mevcc of the nominal $\Bs$ mass are selected, and from them the
invariant mass difference, $\Delta M=M(\Dsp\pim\pip)-M(\Dsp$) is formed, where both $\pip\pim$ combinations are
included.  The $\Delta M$ distribution for candidates in the $\Bsb$ signal window is shown in Fig.~\ref{fig:Bs2Ds1Pi}. 
A peak corresponding to the $D_{s1}(2536)^+$ is observed, whereas no
significant structures are observed in the upper $\Bsb$ mass sideband (5450$-$5590~\mevcc).
The distribution is fitted to the sum of a signal Breit-Wigner shape convolved with a 
Gaussian resolution function, and a second order polynomial to describe the background contribution. 
The Breit-Wigner width is set to 0.92~\mevcc~\cite{Beringer:1900zz}, and the Gaussian resolution
is fixed to 3.8~\mevcc based on simulation.
A signal yield of $20.0\pm5.1$ signal events is observed at a mass difference of $565.1\pm1.0~\mevcc$,
which is consistent with the known $D_{s1}(2536)^+-\Dsp$ mass difference of $566.63\pm0.35~\mevcc$~\cite{Beringer:1900zz}.
The significance of the signal is 5.9, obtained by fitting the invariant mass distribution with the 
mean mass difference fixed to 566.63~\mevcc~\cite{Beringer:1900zz}, and computing $\sqrt{-2{\rm ln}({\mathcal{L}_{0}}/{\mathcal{L}_{\rm max}}})$.
Here, ${\mathcal{L}_{\rm max}}$ and ${\mathcal{L}_{0}}$ are the fit likelihoods 
with the signal yields left free and fixed to zero, respectively. Several variations in the background shape 
were investigated, and in all cases the signal significance exceeded 5.5.
This decay is therefore observed for the first time. To obtain the yield in 
the normalization mode ($\bstodspipipi$), the signal function is integrated from 40~\mevcc below to 40~\mevcc above
the nominal $\Bs$ mass. A yield of $5505\pm85$ events is found in this restricted mass interval.

\begin{figure}[t]
\begin{center}
\includegraphics[width=1.0\linewidth]{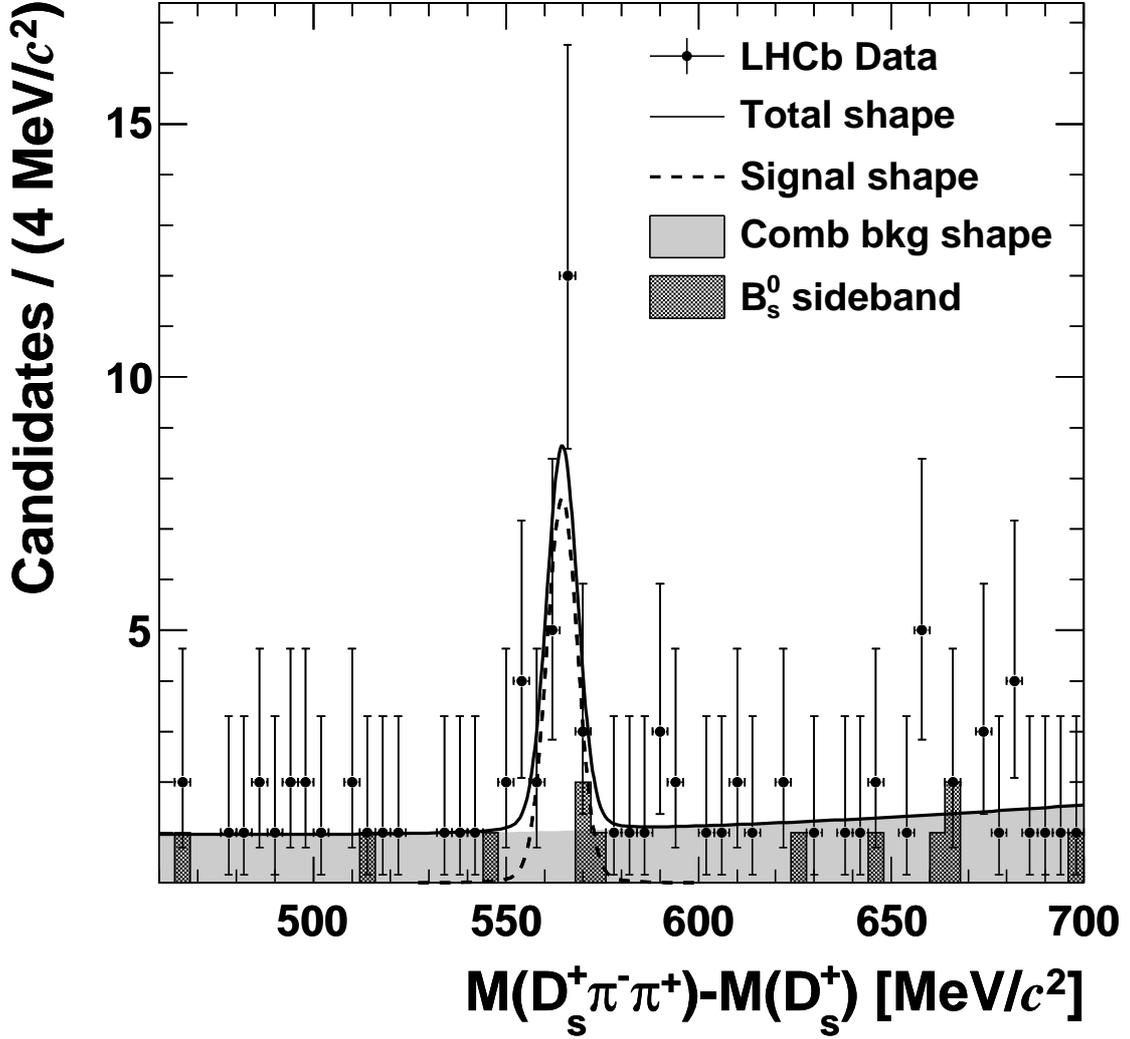}
\end{center}
\caption{Distribution of the difference in invariant mass, $M(\Dsp\pim\pip)-M(\Dsp)$, using
$\bstodspipipi$ candidates within 40~\mevcc of the known $\Bs$ mass (points) and in the upper $\Bs$
mass sidebands (filled histogram). The fit to the distribution is shown, as described in the text.}
\label{fig:Bs2Ds1Pi}
\end{figure}

\section{Selection efficiencies}
\label{sec:effs}

The ratios of branching fractions can be written as
\begin{align}
{\br(\bstodskpipi)\over\br(\bstodspipipi)} &= {Y(\bstodskpipi)\over Y(\bstodspipipi)}\times\erel^s 
\label{eq:bfeq1}
\end{align}
\noindent and
\begin{align}
~~~~~~~~~~~{\br(\btodskpipi)\over\br(\bstodskpipi)} &= {Y(\btodskpipi)\over Y(\bstodskpipi)}\times\erel^d\times f_s/f_d, 
\label{eq:bfeq2}
\end{align}
\noindent where $Y$ are the measured yields, $\erel^{s}=\eff(\bstodspipipi)/\eff(\bstodskpipi)$ and 
$\erel^{d}=\eff(\bstodskpipi))/\eff(\btodskpipi)$ are the relative selection
efficiencies (including trigger), and $f_s/f_d=0.267\pm0.021$~\cite{Aaij:2011jp} is the $\Bs$ fragmentation fraction
relative to $\Bz$. The ratios of selection efficiencies are obtained from
simulation, except for the PID requirements, which are obtained from a dedicated $\Dstarp$ calibration
sample, weighted to match the momentum spectrum of the particles that form $X_d$ and $X_s$. 
The selection efficiencies for each decay are given in Table~\ref{tab:seleff}. The efficiency
of the $\bstodspipipi$ decay is about 35\% larger than the values obtained in either the $\bstodskpipi$ or 
$\btodskpipi$ decay; the efficiencies of the latter two are consistent with each other. The lower efficiency is 
due almost entirely to the tighter PID requirements on the $\Km$ and $\pip$ in $X_s$. Two additional
multiplicative correction factors, also shown in Table~\ref{tab:seleff}, are applied to the measured ratio of 
branching fractions in Eqs.~\ref{eq:bfeq1}~and~\ref{eq:bfeq2}. The first is a correction for the $\Dsp$ mass veto on $M(X_{d,s})$,
and the second is due to the requirement that $M(X_{s,d})<3~\gevcc$. The former, which represents a small correction,
is estimated from the \sWeight-ed distributions of $M(X_{d,s})$ shown previously. For the latter,
the fraction of events with $M(X_{d,s})>3~\gevcc$ is obtained from simulation, and scaled by the
ratio of yields in data relative to simulation for the mass region $2.6<M(X_{s,d})<3.0~\gevcc$.
A 50\% uncertainty is assigned to the estimated correction. Based on the qualitative agreement 
between data and simulation in the $M(X_{d,s})$ distributions (see Sect.~\ref{sec:kinematics}), 
and the fact that the phase space approaches zero as $M(X_{d,s})\to~3.5~\gevcc$, this uncertainty
is conservative.
\begin{table*}[t]
\begin{center}
\caption{Selection efficiencies and correction factors for decay modes under study. The
uncertainties on the selection efficiencies are statistical only, whereas the correction factors
show the total uncertainty.}
\begin{tabular}{lcccc}
\hline\hline
\\[-2.20ex]
Quantity           & $\bstodspipipi$ & $\bstodskpipi$ & $\btodskpipi$ \\
\hline
Total $\eff$ ($10^{-4}$) & $4.97\pm0.08$ & $3.67\pm0.10$ & $3.59\pm0.10$ \\
\hline
$\Ds$ veto corr.  & $1.013\pm0.003$ & $1.013\pm0.003$ & $1.017\pm0.005$ \\
$M>3~\gevcc$ corr. & $1.02\pm0.01$ & $1.04\pm0.02$ & $1.14\pm0.07$ \\
\hline\hline
\end{tabular}
\label{tab:seleff}
\end{center}
\end{table*}
The relative efficiency between $\Bs\to D_{s1}(2536)^+\pim,~D_{s1}^+\to\Dsp\pim\pip$ and $\bstodspipipi$
is estimated from simulation, and is found to be $0.90\pm0.05$.

\section{Systematic uncertainties}

Several uncertainties contribute to the ratio of branching fractions. 
The sources and their values are listed in Table~\ref{tab:syst}.
The largest uncertainty, which applies only to the ratio ${\br(\btodskpipi)\over\br(\bstodskpipi)}$,
is from the $b$ hadronization fraction, $f_s/f_d=0.267\pm0.021$~\cite{Aaij:2011jp}, which is 7.9\%.
Another large uncertainty results from the required correction factor to account for the signal with $M(X_{s,d})>3~\gevcc$. 
Those corrections are described in Sect.~\ref{sec:effs}.

The selection efficiency depends slightly on the modeling of the $X_{d,s}$ decay. The momentum
spectra of the $\Bb$, $\Dsp$, $X_{d,s}$ and the $X_{d,s}$ daughters have been compared to simulation,
and excellent agreement is found. The selection efficiency is consistent with being flat as a function of
$M(X_{d,s})$ at the level of two standard deviations or less. To assess a potential systematic uncertainty due
to a possible $M(X_{d,s})$-dependent efficiency, 
the relative differences between the nominal selection efficiencies and the ones obtained by
reweighting the measured efficiencies by the $X_{d,s}$ mass spectra in data are computed. 
The relative deviations of 0.5\%, 1.1\% and 1.2\% for $\bstodskpipi$, $\bstodspipipi$ and $\btodskpipi$, respectively,
are the assigned uncertainties. The systematic uncertainty on the BDT efficiency is determined by fitting the
$\bstodspipipi$ mass distribution in data with and without the BDT requirement. The efficiency is found to agree with
simulation to better than the 1\% uncertainty assigned to this source. In total, the
simulated efficiencies have uncertainties of 1.6\% and 1.9\% in the two ratios of
branching fractions. The PID efficiency uncertainty is dominated by the usage of the $\Dstarp$ calibration sample
to determine the efficiencies of a given PID requirement~\cite{LHCb-PROC-2011-008}. 
This uncertainty is assessed by comparing the PID efficiencies obtained directly 
from simulated signal decays with the values obtained  using a simulated $\Dstarp$ calibration sample 
that is re-weighted to match the kinematics of 
the signal decay particles. Using this technique, an uncertainty of 2\% each on the $\bstodskpipi$ and 
$\btodskpipi$ PID efficiencies is obtained, which is 100\% correlated, and a 1\% uncertainty for $\bstodspipipi$.
The trigger is fully simulated, and given the identical number of tracks and the well-modeled $\pt$ spectra,
the associated uncertainty cancels to first order. Based on previous studies~\cite{Aaij:2011rj}, 
a 2\% uncertainty is assigned.

The uncertainties in the signal yield determinations have contributions from both the background and signal
modeling. The signal shape uncertainty was estimated by varying all the fixed signal shape parameters one
at a time by one standard deviation, and adding the changes in yield in quadrature (0.5\%). A double Gaussian
signal shape model was also tried, and the difference was negligible.
For the combinatorial background, the shape was modified from a single exponential to either the sum of 
two exponentials, or a linear function. For $\bstodspipipi$, the difference in yield was 0.4\%. For $\bstodskpipi$, 
the maximum change was 4\%, and for $\btodskpipi$, the maximum shift was 1\%. In the 
$\Bzb_{(s)}\to\Dsp\Km\pip\pim$ mass fit, the $\Bzb_{(s)}\to D_s^{*+}\Km\pip\pim$ contribution was 
modeled using the shape from the $\bstodspipipi$ mass fit. To estimate an uncertainty from this assumption, 
the data were fitted with the shape obtained from $\bstodsstarkpipi$ simulation. 
A deviation of 5.5\% in the fitted $\btodskpipi$ yield is found, with almost no change in the $\bstodskpipi$ yield. The 
larger sensitivity on the $\Bzb$ yield than the $\Bsb$ yield arises because these background contributions have a rising edge in the
vicinity of the $\Bzb$ mass peak, which is far enough below the $\Bsb$ mass peak to have negligible impact.
These yield uncertainties are added in quadrature to obtain the values shown in Table~\ref{tab:syst}.
The uncertainties due to the finite simulation sample sizes are 3.0\%.
\begin{table*}[t]
\begin{center}
\caption{Summary of systematic uncertainties (in \%) on the measurements of the ratios of branching fractions.}
\begin{tabular}{lcc}
\hline\hline
\\[-2.20ex]
Source          & ${\br(\bstodskpipi)\over \br(\bstodspipipi)}$ & ${\br(\btodskpipi)\over\br(\bstodskpipi)}$ \\
\\[-2.20ex]
\hline
$f_s/f_d$            &     -      &              7.9   \\
$M(X_{s,d})>3~\gevcc$ &    2.2      &             7.0  \\
Efficiency           &    1.6      &              1.9 \\
PID                  &    2.2      &              0.0  \\
Trigger              &    2.0      &              2.0 \\
Signal yields        &    4.0      &              6.9 \\
Simulated sample size &    3.0      &              3.0 \\
\hline
Total               &     6.4     &               13.4    \\
\hline\hline
\end{tabular}
\label{tab:syst}
\end{center}
\end{table*}

The major source of systematic uncertainty 
on the branching fraction for $\Bsb\to D_{s1}(2536)^+\pim,~\D_{s1}^+\to\Dsp\pim\pip$, 
is from the relative efficiency (5\%), and on the fraction of events with
$M>3~\gevcc$ (10\%). This 10\% uncertainty is conservatively estimated by assuming a flat distribution in $M(X_d)$
up to 3~\gevcc, and then a linear decrease to zero at the phase space limit of $\sim$3.5~\gevcc.
Other systematic uncertainties related to the fit model are negligible. 
Thus in total, a systematic uncertainty of 11\% is assigned to the ratio
$\br(\Bsb\to D_{s1}(2536)^+\pim,~\D_{s1}^+\to\Dsp\pim\pip)/\br(\bstodspipipi)$.

\section{Results and summary}

This paper reports the first observation of the $\bstodskpipi$, $\btodskpipi$ and 
$\Bsb\to D_{s1}(2536)^+\pim,~\D_{s1}^+\to\Dsp\pim\pip$ decays.
The ratios of branching fractions are measured to be
\begin{align*}
{\br(\bstodskpipi)\over\br(\bstodspipipi)} &= (5.2\pm0.5\pm0.3)\times10^{-2} \\
{\br(\btodskpipi)\over\br(\bstodskpipi)} &= 0.54\pm 0.07\pm0.07, 
\end{align*}
\noindent and
\begin{align*}
{\br(\Bsb\to D_{s1}(2536)^+\pim,~D_{s1}^+\to\Dsp\pim\pip)\over\br(\bstodspipipi)} = (4.0\pm1.0\pm0.4)\times10^{-3}, 
\end{align*}
where the uncertainties are statistical and systematic, respectively.
The $\bstodskpipi$ branching fraction is consistent with expectations from Cabibbo suppression.
This decay is particularly interesting because it can be used in a time-dependent analysis to 
measure the CKM phase $\gamma$. Additional
studies indicate that this decay mode, with selections optimized for only $\bstodskpipi$, can 
contribute about an additional 35\% more signal events relative to the signal yield in $\bstodsk$ alone.

The $\btodskpipi$ branching fraction is about 50\% of that for  $\bstodskpipi$. Compared to
the $\btodsk$ decay that proceeds only via a $W$-exchange diagram, where $\br(\btodsk)/\br(\bstodskcc)\sim0.1$~\cite{Beringer:1900zz}, 
the ratio $\br(\btodskpipi)/\br(\bstodskpipi)$ is about 5 times larger.  A 
consistent explanation of this larger $\btodskpipi$ branching fraction is that only about
1/5 of the rate is from the $W$-exchange process (Fig.~\ref{fig:feyn}(d))
and about 4/5 comes from the diagrams shown in Figs.~\ref{fig:feyn}(e-f).
The observed $M(X_s)$, $M(\Km\pip)$ and $M(\pip\pim)$ distributions in Fig.~\ref{fig:Bd2DsK2Pi_Masses} 
also support this explanation, as evidenced by the qualitative agreement with the simulation.

\section*{Acknowledgements}

\noindent We express our gratitude to our colleagues in the CERN
accelerator departments for the excellent performance of the LHC. We
thank the technical and administrative staff at the LHCb
institutes. We acknowledge support from CERN and from the national
agencies: CAPES, CNPq, FAPERJ and FINEP (Brazil); NSFC (China);
CNRS/IN2P3 and Region Auvergne (France); BMBF, DFG, HGF and MPG
(Germany); SFI (Ireland); INFN (Italy); FOM and NWO (The Netherlands);
SCSR (Poland); ANCS/IFA (Romania); MinES, Rosatom, RFBR and NRC
``Kurchatov Institute'' (Russia); MinECo, XuntaGal and GENCAT (Spain);
SNSF and SER (Switzerland); NAS Ukraine (Ukraine); STFC (United
Kingdom); NSF (USA). We also acknowledge the support received from the
ERC under FP7. The Tier1 computing centres are supported by IN2P3
(France), KIT and BMBF (Germany), INFN (Italy), NWO and SURF (The
Netherlands), CIEMAT, IFAE and UAB (Spain), GridPP (United
Kingdom). We are thankful for the computing resources put at our
disposal by Yandex LLC (Russia), as well as to the communities behind
the multiple open source software packages that we depend on.



\addcontentsline{toc}{section}{References}
\bibliographystyle{LHCb}
\bibliography{main}

\end{document}